\documentclass[runningheads]{llncs}
\usepackage[T1]{fontenc}
\usepackage{algorithm}
\usepackage{algorithmic}
\usepackage{amsmath}
\usepackage{todonotes}
\usepackage{graphicx}
\usepackage{comment}  

\usepackage{tabularx}
\usepackage{array}
\usepackage{booktabs}
\usepackage{adjustbox}
\usepackage{multirow}
\usepackage{xcolor, colortbl}
\usepackage{dirtytalk}
\usepackage{hyperref}
\usepackage{amssymb}

\definecolor{cgreen}{rgb}{0.2,0.6,1}
\definecolor{cred}{rgb}{0.968,0.545,0.321}
\definecolor{brightpink}{rgb}{1.0, 0.0, 0.5}

\newcommand{\rtr}[1]{{\scriptsize\color{cgreen}$\blacktriangledown$ #1}}

\newcommand{\gtr}[1]{{\scriptsize\color{brightpink}$\blacktriangle$ #1}}
\usepackage{hyperref}
\hypersetup{
    colorlinks=true,
    linkcolor=blue,
    filecolor=magenta,    
}
\begin{document}
\title{MedEdit: Counterfactual Diffusion-based Image Editing on Brain MRI}
\vspace{-5em}
%
%

\author{Malek Ben Alaya\inst{1} \and
Daniel M. Lang\inst{1,2} \and
Benedikt Wiestler\inst{1,3} \and Julia A. Schnabel$^*$ \inst{1,2,4} \and Cosmin I. Bercea$^*$ \inst{1,2}}
%
%
\institute{Technical University of Munich, Munich, Germany \and
Helmholtz AI and Helmholtz Center Munich, Munich, Germany \and
Klinikum Rechts der Isar, Munich, Germany \and
Kings College London, London, UK }

%
\authorrunning{M.Ben Alaya, D.Lang et al.}
%

%
\maketitle              
\vspace{-2.7em}

\begin{abstract}
Denoising diffusion probabilistic models enable high-fidelity image synthesis and editing. In biomedicine, these models facilitate counterfactual image editing, producing pairs of images where one is edited to simulate hypothetical conditions. For example, they can model the progression of specific diseases, such as stroke lesions. 
However, current image editing techniques often fail to generate realistic biomedical counterfactuals, either by inadequately modeling indirect pathological effects like brain atrophy or by excessively altering the scan, which disrupts correspondence to the original images. Here, we propose \emph{MedEdit}, a conditional diffusion model for medical image editing. \emph{MedEdit} induces pathology in specific areas while balancing the modeling of disease effects and preserving the original scan's integrity. We evaluated \emph{MedEdit} on the Atlas v2.0 stroke dataset using Frechet Inception Distance and Dice scores, outperforming state-of-the-art diffusion-based methods such as Palette (by 45\%) and SDEdit (by 61\%). Additionally, clinical evaluations by a board-certified neuroradiologist confirmed that \emph{MedEdit} generated realistic stroke scans indistinguishable from real ones. We believe this work will enable counterfactual image editing research to further advance the development of realistic and clinically useful imaging tools.

\vspace{-2.3em}
\begin{figure*}[htb!]
\centering
      \includegraphics[width=0.73\linewidth]{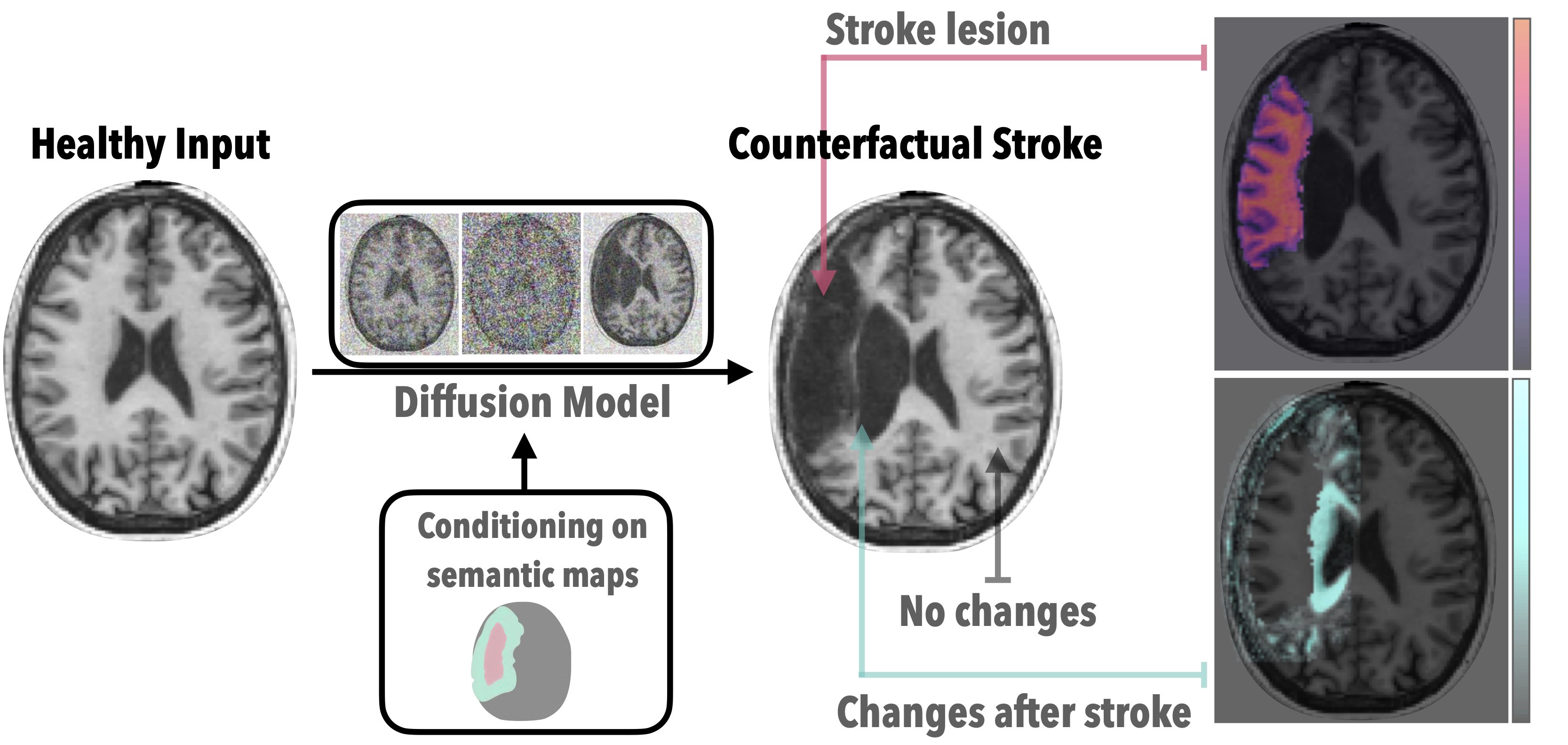} 
    \caption{Overview of \emph{MedEdit}. It conditionally edits prior scans to generate counterfactual stroke scans that simulate direct and indirect pathological effects.}
   \end{figure*}
\end{abstract}
\vspace{-5em}

\def\thefootnote{*}\footnotetext{Equal Contribution}\def\thefootnote{\arabic{footnote}}

\section{Introduction}
Counterfactuals involve exploring \say{what if} scenarios to investigate potential outcomes of different interventions in patient conditions. In the medical context, these hypothetical constructs enable researchers and clinicians to gain insights into causal relationships and underlying mechanisms of disease progression. From a predictive modeling perspective, generated counterfactuals can be used for various applications, such as data augmentation to enhance machine learning models on underrepresented populations~\cite{shin2018medical}, stress-testing models for population or acquisition shifts to uncover biases~\cite{pérezgarcía2024radedit}, and providing counterfactual explanations to understand the decision mechanism of classification models~\cite{atad2022chexplaining}.

Recent studies have focused on biomedical counterfactual image editing, particularly on chest X-ray datasets~\cite{gu2023biomedjourney,pérezgarcía2024radedit}. These datasets are widely available, include paired scans showing disease progression, and are multimodal, incorporating text descriptions alongside images.
In contrast, brain imaging studies on counterfactual editing predominantly focus on generating healthy scans from pathological ones to aid in lesion localization~\cite{bercea2023mask,sanchez2022healthy,wyatt2022anoddpm}. Some studies tackle counterfactual generation for Alzheimer’s disease progression using paired images~\cite{ad_diffusion,puglisi2024enhancing}. However, counterfactual disease editing for brain imaging from \mbox{unpaired} images remains largely unexplored.

Several unpaired image-to-image translation methods based on diffusion models have been developed for the natural image domain. They have been used on datasets like ImageNet~\cite{deng2009imagenet} and Places2~\cite{zhou2017places}. SDEdit~\cite{meng2022sdedit} is an image editing technique initially designed to turn sketches into realistic images. The editing process starts by diffusing the image with Gaussian noise up to a specific timestep, then denoising it using a diffusion network. Palette~\cite{saharia2022palette} is an inpainting method, which trains a diffusion model to fill in the missing parts of an image using the known regions as a condition. 
To the best of our knowledge, such diffusion-based methods have not yet been adopted in the medical domain.

In this work, we propose a conditional diffusion-based image editing approach that generates realistic counterfactual pathological brain scans, using unpaired data during model training. Unlike competing methods, our method can model indirect pathological changes that can be caused by a pathology (e.g. brain atrophy caused by a stroke) while having a high fidelity to the prior factual scan. We believe that our proposed study will open new avenues for exploring counterfactual biomedical image editing on brain imaging for new modalities and new pathology types. We summarize our main contributions below:
\begin{itemize}
    \item We benchmark state-of-the-art image editing and inpainting methods on generating realistic counterfactual brain scans with stroke lesions.
    \item We propose \emph{MedEdit}, a novel biomedical counterfactual image editing method that can simulate diseases and model their realistic consequences, while maintaining high fidelity to the prior scan.
    \item We validate our findings through anonymized clinical tests conducted by a board-certified neuroradiologist, assessing realism, fidelity to the prior scan, and the accurate modeling of pathological changes, including induced diseases and their realistic effects.
\end{itemize}

\section{Background}
{\textbf{Denoising Diffusion Probablistic Models }(DDPMs)~\cite{ho2020denoising} are a class of generative models that enable sampling from data distributions by learning to denoise samples that have been corrupted by Gaussian noise. DDPMs operate by establishing a forward-time process that incrementally adds noise to original samples \( x_{0} \) for \( t = 1, \ldots, T \) through:

\begin{equation}
q\left(x_t \mid x_{t-1}\right) \sim \mathcal{N}\left(x_t ; \sqrt{1 - \beta_{t}}x_{t-1},\beta_t \mathbf{I}\right),
\end{equation}
where the noise schedule \( \beta_{1:T} \) defines the level of noise added and is an increasing function of the timestep $t$, ensuring that \( x_{T} \) is (almost) pure Gaussian noise. Using the independence property of the noise added at each step of (1), we get:
\begin{equation}
    q\left(x_t \mid x_0\right) \sim \mathcal{N}\left(x_t ; \sqrt{\bar{\alpha}_{t}} \, x_{0},\sqrt{1 - \bar{\alpha}_{t}}  \mathbf{I}\right).
\end{equation}
This can be written as:
\begin{equation}
    x_{t} = \sqrt{\bar{\alpha}_{t}} \, x_{0} + \sqrt{1 - \bar{\alpha}_{t}} \, \bar{\epsilon}_{t},
\end{equation}
where \( \bar{\epsilon}_{1:T} \sim \mathcal{N}(0, I) \) and \( \bar{\alpha}_{t} = \prod_{s=1}^{t} (1 - \beta_{s}) \). To synthesize new images, Gaussian noise is reversed back into samples from the learned distribution. Although the exact reversal of the forward process is intractable, a variational approximation is achieved by minimizing the denoising objective~\cite{ho2020denoising} at training time:

\begin{equation}
    \mathcal{L} = \mathbb{E}_{\mathbf{x}_{0}, t, \boldsymbol{\epsilon}} \left\| \boldsymbol{\epsilon} - \epsilon_{\theta} \left( \mathbf{x}_{t}, t \right) \right\|_{2}^{2}.
\end{equation}
The variational approximation is defined through the following equations:

\begin{equation}
    x_{t-1} = \hat{\mu}_{t} \left( x_{t}, t \right) + \sigma_{t} z_{t},
\end{equation}

\begin{equation}
    \hat{\mu}_{t} \left( x_{t}, t \right) = \frac{1}{\sqrt{\alpha_{t}}} \left( \mathbf{x}_{t} - \frac{\beta_{t}}{\sqrt{1 - \bar{\alpha}_{t}}} \boldsymbol{\epsilon}_{\theta} \left( \mathbf{x}_{t}, t \right) \right),
\end{equation}
where \( \epsilon_{\theta} \left( x_{t}, t \right) \) is a learned approximation of the noise \( \bar{\epsilon}_{t} \) that corrupted the original image \( x_{0} \) to produce \( x_{t} \), which can be parameterized with a U-Net architecture~\cite{ho2020denoising,ronneberger2015u}. Here, \( z_{1:T} \sim \mathcal{N}(0, I) \) and \( \sigma_{1:T} \) defines the level of noise introduced.  For $\sigma_{t} = 0$, the process is deterministic and is referred to as a Denoising Diffusion Implicit Model (DDIM)~\cite{song2022denoising}. For probabilistic models with $ \sigma_{t} = \sqrt{\frac{1-\bar{\alpha}_{t-1}}{1-\bar{\alpha}_{t}} \beta_{t}} $, the process is known as a DDPM [27]. We use such value for $\sigma_{t}$ throughout this work.\\


{\noindent\textbf{Learning conditional distributions with DDPMs}} involves modifying the denoiser network $\epsilon_\theta$ to take a conditioning signal as input, e.g. a bounding box, a text prompt, or a semantic map. Such conditioning can be achieved through cross-attention or simple concatenation of the signal to the input channels of the denoiser network $\epsilon_\theta$~\cite{rombach2022highresolution}.This translates to setting the conditional signal $c$ in the set of equations above, thus changing $\boldsymbol{\epsilon}_{\theta} \left( \mathbf{x}_{t}, t \right)$ to $\boldsymbol{\epsilon}_{\theta} \left( \mathbf{x}_{t}, c, t \right)$.


\noindent{\textbf{Generic image editing with RePaint.}}
 Re-paint~\cite{lugmayr2022repaint} adapts the reverse process of diffusion models to enable the inpainting of specific areas by sampling from a joint distribution of a learned set of images. 
The method has two key components. First the inpainting of the unknown regions is conditioned on the known regions. This ensures that the unpainted region shares meaningful semantics with the unmasked area. However, the two areas might still show inconsistencies. The second component, known as resampling, addresses this issue. It harmonizes the two different regions by repeating the conditioning process. More specifically, this is done by diffusing $x_{t-1}$ back to $x_t$ and reapplying the conditioning process. This technique is known as \say{resampling steps} and can be done multiple times.

\section{Method}
We introduce \emph{MedEdit}, a conditional diffusion-based counterfactual image editing algorithm tailored to balance the modeling of indirect pathological changes with high fidelity to the original scan during pathology simulation. \emph{MedEdit} extends the original RePaint algorithm by converting its class-conditional inpainting process into a mask-conditioned one, enabling targeted pathology simulation. Additionally, it introduces a mask selection method of the region to be inpainted facilitating the representation of potential indirect pathological changes.\\

\noindent{\textbf{Conditional inpainting.}} 
We introduce the conditions to the diffusion model by concatenating the masks as additional input channels to the denoiser network of the diffusion model. This is shown in line \ref{line:7_} of Algorithm \ref{alg:mededit}: 
\begin{equation}
    x_{t-1}^{\text{unknown}} = \frac{1}{\sqrt{\alpha_t}}\left(x_t - \frac{\beta_t}{\sqrt{1-\bar{\alpha}_t}}\epsilon_\theta(x_t, b, p, t)\right) + \sigma_tz,
\end{equation}
where $b$ and $p$ are the brain and the pathology masks, respectively.\\

\noindent{\textbf{Mask selection.}}
We select the mask of the region to be inpainted such that indirect pathological changes can be modeled during the editing process. 
Naïvely choosing $m = p$  would only inpaint a pathology in the desired area, without accounting for the changes it may cause in other areas of the brain. We introduce this method, which we call naïve RePaint, as a baseline in our experiments. To model the required indirect pathological changes, we choose $m$ to be a diluted version of the desired pathology mask $p$. The dilution kernel size $k$ controls the positional extent of the indirect pathological changes. A detailed description of \emph{MedEdit} is provided in Algorithm \ref{alg:mededit}.

\begin{algorithm}[tb]
\caption{Biomedical counterfactual image editing with MedEdit.}
\label{alg:mededit}
\begin{algorithmic}[1]
\STATE $x_T \sim \mathcal{N}(0, I), x_0: \text{prior scan}, b: \text{brain mask}, p: \text{pathology mask}$
\STATE $k: \text{kernel size}$
\STATE $m = dilute(p,k)$
\FOR{$t = T, \ldots, 1$}
    \FOR{$u = 1, \ldots, U$}
        \STATE $\epsilon \sim \mathcal{N}(0, I)$ \textbf{if} {$t > 1$} \textbf{else} $\epsilon = \mathbf{0}$ \label{line:4_}
        
        \STATE $x_{t-1}^{\text{known}} = \sqrt{\bar{\alpha}_t}x_0 + (1 - \bar{\alpha}_t)\epsilon$

        \vspace{0.5em}  
        
        \STATE $z \sim \mathcal{N}(0, I)$ \textbf{if} {$t > 1$} \textbf{else} $z = \mathbf{0}$
        \STATE $x_{t-1}^{\text{unknown}} = \frac{1}{\sqrt{\alpha_t}}\left(x_t - \frac{\beta_t}{\sqrt{1-\bar{\alpha}_t}}\epsilon_\theta(x_t, b, p, t)\right) + \sigma_tz$ \label{line:7_}
        
        \vspace{0.5em}  
        
        \STATE $x_{t-1} = m \odot x_{t-1}^{\text{known}} + (1 - m) \odot x_{t-1}^{\text{unknown}}$ \label{line:8_}

        \vspace{0.5em}  
        
        \IF{$u < U$ \textbf{and} $t > 1$} \label{line:9_}
            \STATE $x_t \sim \mathcal{N}(\sqrt{1 - \beta_{t-1}}x_{t-1}, \beta_{t-1}I)$ \label{line:10_}
        \ENDIF
    \ENDFOR
\ENDFOR
\STATE \textbf{return} edited version of $x_0$
\end{algorithmic}
\end{algorithm}

\section{Experiments}
We conduct a comprehensive set of experiments to benchmark the performance of our proposed method against state-of-the-art image editing and inpainting methods in simulating stroke effects on brain scans. Our evaluation focused on realism, adherence to desired pathological change, fidelity to the prior scan, and modeling of indirect pathological changes like brain atrophy, using clinical and computational metrics such as FID and Dice scores. We compare our proposed method to Palette~\cite{saharia2022palette}, SDEdit~\cite{meng2022sdedit} and naïve RePaint. We adapt  SDEdit to denoise conditionally on brain and pathology masks, similarly to how Couairon et al.~\cite{couairon2022diffedit} adapted the model to perform denoising conditioned on text prompts.\\

\noindent{\textbf{Dataset.}}
We use the Atlas v2.0 dataset~\cite{liew2022large}, which contains 655 T1-w brain Magnetic Resonance Imaging scans. We normalize the mid-axial slices to the 98th percentile, apply padding, and resize them to a resolution of 128 × 128. Of the total 655 images, only 443 contain a pathology. We stratify the pathological subset with respect to the pathology size into three pathology groups, namely small, medium and large. The small group (N=111) comprises the first 25th percentile, consisting of lesions smaller than 18.5 pixels. The large group (N=111) encompasses the top 25th percentile, including lesions larger than 371 pixels. The medium group (N=221) includes the remaining scans with lesions of intermediate sizes. We further split the pathological subset into a train (N=389) and test set (N=54).\\

\noindent{{\textbf{Implementation details.}}} We train a diffusion model to generate pathological brain scans, conditioned on brain and pathology masks. This model is later used to generate counterfactuals for SDEdit, naïve RePaint and MedEdit. The U-Net from~\cite{ho2020denoising} is utilised, along with $T = 1000$ and a linear noise coefficient $\beta_t$ ranging from $ \beta_1 = 10^{-4}$ to $\beta_T = 0.02$ as in~\cite{ho2020denoising}. The training lasts for 1500 epochs. In \emph{MedEdit}, we use $k=25$ with four resampling steps. For naive RePaint, three resampling steps are used. For SDEdit, we use an encoding ratio of $0.2$.\\

\noindent{{\textbf{Evaluation.}}}
At test time, we generate counterfactuals by randomly pairing each pathology mask from the test set with a scan from the set that doesn't contain a pathology, resulting in triplets of (prior, brain mask of prior, pathology mask).

\noindent\textit{Computational metrics.} We assess the realism of the generated counterfactuals by computing the Frechet Inception Distance (FID)~\cite{fid} to the real pathological test distribution. We used nnUNet~\cite{isensee2021nnu} to identify the pathology lesions in the generated counterfactuals. We compute Dice scores by measuring the overlap to the ground truth pathology masks to evaluate the adherence to desired pathological changes. We compute these metrics over 10 bootstrapping runs. 

\noindent\textit{{Clinical metrics:}}
We provide the generated counterfactuals to a board-certified neuroradiologist for clinical assessment to rate the validity of the computational metrics utilized and further assess the fidelity of the generated counterfactuals relative to their original scans and whether they account for indirect pathological changes. First, we randomly select 20 counterfactuals from each of the benchmarked methods, stratified by pathology size. To assess their realism, we mix these with 20 real samples from the test set. We then pass this combined set of scans (N=100) to the first part of the clinical assessment. Here, the realism of the counterfactuals is rated on a scale from 1 to 5. For the second part, we use the same counterfactuals alongside the corresponding prior scans. Here, adherence to desired pathological change (Path.), fidelity to the original scan (Fidel.), and whether indirect pathological changes are accurately modeled (Ind-Path.) are rated on a scale from 1 to 5. We henceforth refer to the ratings of the first and second part of the clinical assessment as clinical metrics. 

\begin{table}[t]
\caption{We evaluated the perceived realism, fidelity to the prior scan (Fidel.), adherence to desired pathological change (Path.), and modeling indirect pathological changes (Ind-Path.). Best results are shown in \textbf{bold} and second-best are \underline{underlined}. \gtr{}\rtr{} show performance changes relative to the best method.}
\label{table:comparison}
\centering
\small 
\begin{adjustbox}{width=0.9\linewidth,center} 
    \begin{tabular}{l | c | c c || c c c c}
        \toprule
        \multirow{2}{*}{Method} & \multicolumn{3}{c||}{Computational Metrics} & \multicolumn{4}{c}{Clinical Metrics} \\
        & (1-Dice) * Fid $\downarrow$ & Fid $\downarrow$ & Dice $\uparrow$ &  Realism $\uparrow$ & Fidel. $\uparrow$ & Path. $\uparrow$ & Ind-Path. $\uparrow$ \\
        \midrule
        Real samples & - & - & - & \textbf{3.20} & - & - & - \\
        SDEdit~\cite{meng2022sdedit} & 7.95~\gtr{159\%} & 24.1 & \textbf{0.67} & \underline{2.80} & 2.10 & 3.60 & \underline{3.00} \\
        Palette~\cite{saharia2022palette} & 5.63~\gtr{83\%} & 9.08 & 0.38 & 2.40 & \underline{3.95} & \underline{3.65} & 2.00 \\
        Naïve RePaint & \underline{4.24}~\gtr{38\%} & \underline{8.31} & 0.5 & 2.55 & \textbf{4.00} & \textbf{3.70} & 1.85 \\ \midrule
        \rowcolor{gray!10} \emph{MedEdit} (ours) & \textbf{3.07}~\rtr{28\%} & \textbf{8.30} & \underline{0.63} & \textbf{3.20} & 3.20 & 3.45 & \textbf{3.15} \\
        \bottomrule
    \end{tabular}
\end{adjustbox}
\end{table}

\begin{figure*}[ht!]
\centering
      \includegraphics[width=0.85\linewidth]{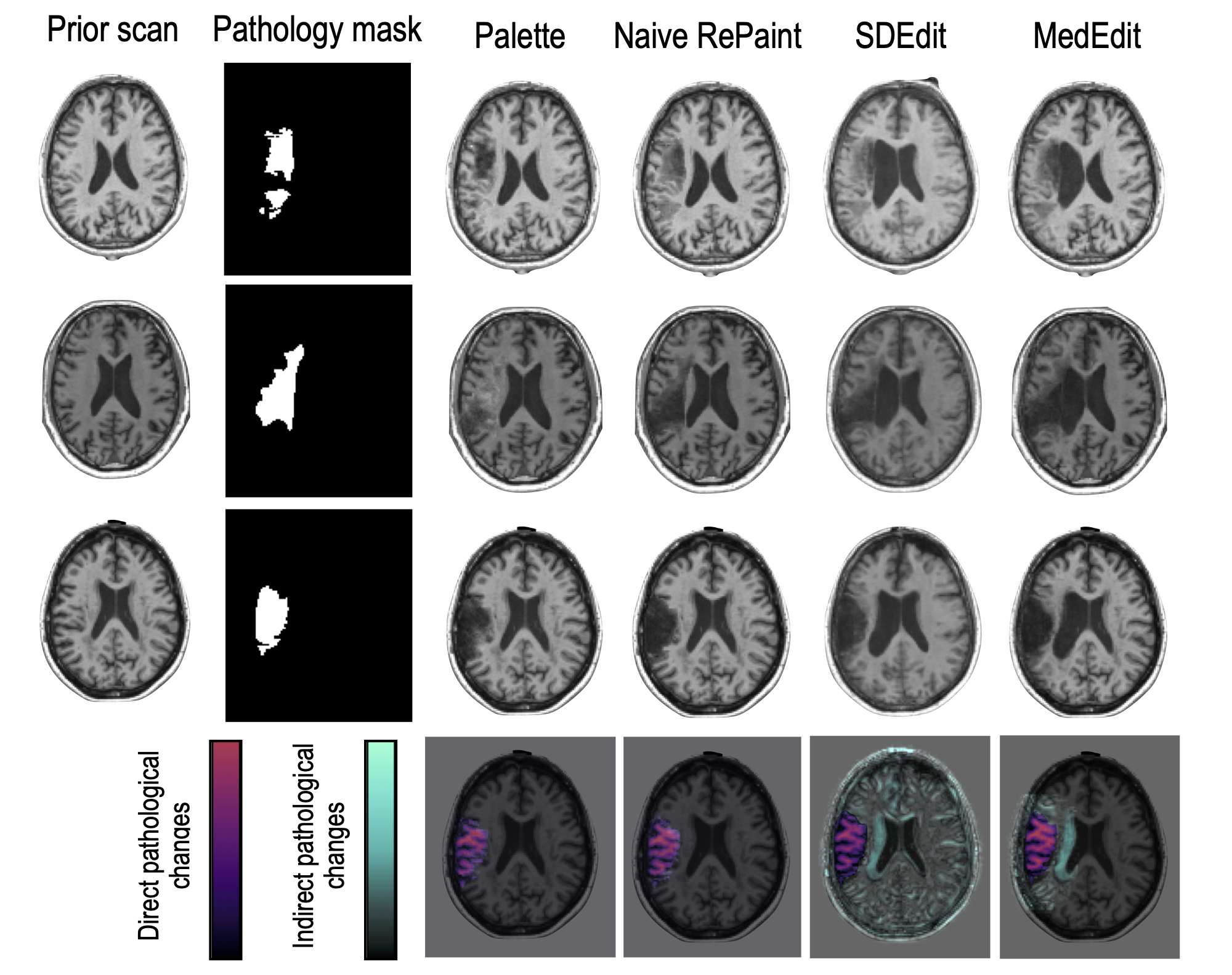} 
    \caption{Examples of counterfactuals obtained with Palette, Naïve RePaint, SDEdit and \emph{MedEdit}. All methods model the pathology well for the last case in the bottom row (shown in purple difference maps). Additionaly, \emph{MedEdit} also precisely models indirect pathological changes induced by the pathology, as shown in turquoise. In this case the stroke lesions caused the ventricle on the same side to enlarge.\label{fig::quali}}
   \end{figure*}

\section{Results}
Quantitative and qualitative results are presented in Table~\ref{table:comparison} and Figure~\ref{fig::quali}. From a computational metrics perspective, \emph{MedEdit} achieves the lowest FID score, indicating superior alignment with the distribution of real stroke images compared to baseline methods. Notably, \emph{MedEdit} outperforms SDEdit by approximately 65.6\%. In the downstream lesion segmentation evaluation, \emph{MedEdit} substantially outperforms Palette and naïve RePaint, with improvements of approximately 65.8\% and 26\%, respectively, while closely trailing SDEdit, with its performance only marginally lower ($\approx$ 6\%). Nevertheless, \emph{MedEdit} ranks highest in a balanced evaluation based on the combined (1-Dice) * FID metric.

From a clinical metrics perspective, \emph{MedEdit} achieves a realism level comparable to that of real samples, providing a 14\% more realistic synthesis than SDEdit, 25\% more than naïve RePaint, and 33\% more than Palette. Although all methods show similar clinical adherence to desired pathological changes, \emph{MedEdit} scores slightly lower. Moreover, our analysis reveals a trade-off in modeling indirect pathological changes versus preserving healthy brain features. Naïve RePaint and Palette, while preserving healthy features, fail to adequately model indirect pathological changes, as shown in Figure~\ref{fig::quali}. In contrast, \emph{MedEdit} and SDEdit manage this balance more effectively, with \emph{MedEdit} outperforming SDEdit by providing a better preservation of healthy brain tissues, while achieving comparable modeling of indirect pathological changes. This balance is demonstrated in the visual comparisons in Figure~\ref{fig::quali}.

\section{Discussion}

The computational metrics currently used, such as the Fréchet Inception Distance (FID) and Dice coefficient, though useful for basic comparisons, fall short in capturing nuanced clinical realities essential in medical imaging. FID, for instance, assesses general image distribution alignment, but overlooks critical subtleties like the indirect pathological effects accompanying stroke lesions, which are vital for a comprehensive clinical evaluation. This discrepancy is clearly demonstrated in Table~\ref{table:comparison}, where, despite similar FID scores for Naïve RePaint and \emph{MedEdit}, their clinical evaluations differ markedly. \emph{MedEdit} excels in capturing indirect pathological changes, showing a 70\% improvement over Naïve RePaint. Similarly, the Dice coefficient focuses narrowly on lesion segmentation accuracy, disregarding other realistic attributes such as edema or secondary tissue changes that are clinically significant. This discrepancy between computational assessments and clinical relevance points to an urgent need for more sophisticated metrics that can holistically evaluate both the primary and secondary effects of pathological conditions in a manner that aligns with clinical observations and patient outcomes.

\emph{MedEdit} facilitates the generation of counterfactual images depicting stroke lesions, which is vital for paired healthy-diseased medical image analysis. This capability enhances understanding of disease progression and improves data augmentation strategies, which are crucial for increasing diagnostic accuracy, training medical professionals, and supporting personalized treatment planning. Future work could extend its applications to three-dimensional imaging and include modeling of global indirect pathological changes. We did not design \emph{MedEdit} specifically for stroke synthesis in brain imaging and believe its methodology holds potential for adaptation to other diseases and organs, significantly expanding its applicability across various medical fields.

\section{Conclusion}
In conclusion, our study adresses the challenge of counterfactual image editing for brain scans. We assess existing image editing and inpainting techniques, identifying their limitations in balancing the modeling of indirect pathological changes with the preservation of healthy regions in the original scan. To address these limitations, we introduced \emph{MedEdit}, a novel method that effectively captures this balance, outperforming state-of-the-art diffusion-based image editing methods. Additionally, our findings highlight discrepencies between computational and clinical metrics, underscoring the need for the development of clinically-relevant metrics that allow automated evaluation of generated counterfactuals.

\newpage
\section{Acknowledgments} C.I.B. is funded via the EVUK program (“Next-generation Al for Integrated Diagnostics”) of the Free State of Bavaria and partially supported by the Helmholtz Association under the joint research school \mbox{‘Munich School for Data Science’.}
\bibliographystyle{splncs04}
\bibliography{main}

\end{document}